\documentstyle[multicol,prl,aps]{revtex}
\input epsf
\input{epsf.sty}
\begin{document}

\title{$^{11}$B NMR detection of the magnetic field distribution in the mixed
superconducting state of MgB$_2$.  }
\author{G. Papavassiliou, M. Pissas, M. Fardis, M. Karayanni, and C. Christides \protect\cite{address} }
\address{Institute of Materials Science, National Center for Scientific Research "Demokritos", 153 10 Athens, Greece}
\date{\today }
\maketitle

\begin{abstract}
The temperature dependence of the magnetic field distribution in the mixed
superconducting phase of randomly oriented MgB$_2$ powder was probed by $%
^{11}$B NMR spectroscopy. Below the temperature of the second critical ($B_{%
{\rm c2}}$) field, $T_{{\rm c2}}\approx 27$K, our spectra reveal two NMR
signal components, one mapping the magnetic field distribution in the mixed
superconducting state and the other one arising from the normal state. The
complementary use of bulk magnetization and NMR measurements reveals that 
MgB$_2$ is an anisotropic superconductor with a $B_{c2}^c<2.35$ Tesla anisotropy 
parameter $\gamma\approx 6$.
\end{abstract}


 \begin{multicols}{2}

The recent discovery of superconductivity \cite{Akamitsu01} in MgB$_2$ has
revived the excitement on this area of research because this alloy becomes
superconducting at unexpectedly \cite{Geballe} high temperatures ($T_c=39$
K), for ''light'' main group elements residing between Be and S in the
periodic table. Subsequent studies have shown that MgB$_2$ is a type-II
superconductor with B$_{{\rm c1}}$($0$)$\approx 0.26$ Tesla, B$_{{\rm c2}}$($%
0$)$\approx 14$ Tesla, with a small condensation energy (relative to Nb$_3$%
Sn and YBa$_2$Cu$_3$O$_7$), a $\xi _0$$\approx 4.9$ nm, and a $\lambda _0$$%
\approx 185$ nm \cite{Wang01}. However, opinions about the nature of the
superconductivity mechanism are still contradictory. Band structure
calculations \cite{Kortus01} suggest that MgB$_2$ is a BCS superconductor,
where superconductivity results from in-plane electron-phonon coupling on
the boron sublattice. The detected isotope effect \cite{Bud'ko01} and the
BCS-type energy gap, that is obtained by tunneling spectroscopy \cite
{Bollinger01} and $^{11}$B NMR \cite{Kotegawa01,Gerashenko01,Tou01}, are in
support of this model. However, specific heat measurements indicate that the
superconducting gap is either anisotropic or two-band-like \cite{Wang01}.
Further deviations from the $s$-wave model have been detected on the
temperature dependence of $B_{{\rm c1}}$ and $\lambda $ \cite{Li01}
as well. 

Recently few works\cite{Lima01,0105271,0105545,Simon01,0106577,Patnaik}
presented convincing evidence that the MgB$_2$
is an anisotropic superconductor with an anisotropy parameter 
$\gamma=(B_{{\rm c2}}^{{\rm ab}}/B_{{\rm c2}}^c)$ taking values into interval $2\leq\gamma \leq 6$.

In principle, a strong anisotropy in the mixed superconducting state of
powder MgB$_2$ - if present- should be detectable with $^{11}$B NMR
spectroscopy. As known, for fields $B_{c1}<B_0<B_{c2}$ a vortex lattice is
formed that gives rise to a characteristic magnetic field distribution with
van Hove singularities at fields where $\nabla \cdot B=0$. For a perfect
hexagonal vortex lattice the field distribution exhibits a peak at a value $%
B_s$, which corresponds to the saddle point located midway between two
vortices, whereas two steps at the maximum ($B_{{\rm max}}$) and minimum ($%
B_{{\rm min}}$) fields are expected \cite{Fite66,Brandt}. Such a magnetic
field distribution should be mapped on the NMR line shape, as the Larmor
frequency of the resonating nuclei depends linearly on the local magnetic
field. Succesfull mapping of the magnetic field distribution has been
already presented in a variety of superconducting materials, like vanadium 
\cite{Fite66}, and rare earth nickel borocarbides YNi$_2$B$_2$C with $^{11}$%
B NMR \cite{Lee00}. In case of strong anisotropy and in applied field $B_{%
{\rm c2}}^c<B_0<B_{{\rm c2}}^{ab}$, a powder superconducting sample with
randomly oriented grains is expected to give a superposition of magnetic
field distributions, ranging in between the normal state and the Abrikosov
lattice. It is thus of particular interest to testify the possibility of
anisotropy in the mixed superconducting state of MgB$_2$, by performing 
$^{11}$B NMR experiments in magnetic fields fullfilling the above condition.

In this letter we report $^{11}$B NMR line shape measurements on powder MgB$%
_2$ samples in external magnetic fields $B_0=2.35$ and $4.7$ Tesla, which
exhibit a strong asymmetric low frequency broadening at temperatures lower
than the temperature of the second critical field ($T$$<$$T_{{\rm c2}}$).
The low temperature spectra may be decomposed in two components: One
component corresponding to the unshifted NMR signal of the normal state,
which indicates that a portion of the sample volume remains in the normal
state even at the lowest measured temperature of $5$ K. A second component,
which maps the magnetic field distribution of the vortex lattice in the
mixed state (vortex component). Our data
provide clear evidence that for a part of the MgB$_2$ grains the 
B$_{{\rm c2}}$($0$)$<2.35$ Tesla, whereas for another part of the grains 
B$_{{\rm c2}}$($0$) is sufficiently higher, and according to magnetic and 
CESR measurements \cite{Simon01} B$_{{\rm c2}}^{ab}$(0)$\approx 14$ 
Tesla. Thus, our results provide a strong experimental evidence that 
MgB$_2$ is an anisotropic superconductor with an anisotropy parameter 
$\gamma\approx 6$.

High quality MgB$_2$ powder samples were prepared by liquid-vapor to solid
reaction in an alumina crucible placed inside a vacuum sealed silica tube, 
using a 3\% excess of Mg. Pure Mg and B powders were thoroughly mixed and 
subsequently slowly heated up to $910^{\rm o}$ C. At this temperature the 
sample annealed for two hours and then cooled slowly down to room 
temperature. Rietveld refinement of x-ray powder diffraction spectra 
revealed that the examined sample consists of 95\% MgB$_2$, with cell 
constants equal to $a=b=3.0849(1)$ \AA\ and $c=3.5213(1)$ \AA, and a 
secondary phase of 5\% MgO.
\begin{figure}[tbp] \centering
\centerline{\epsfxsize 7cm \epsfbox{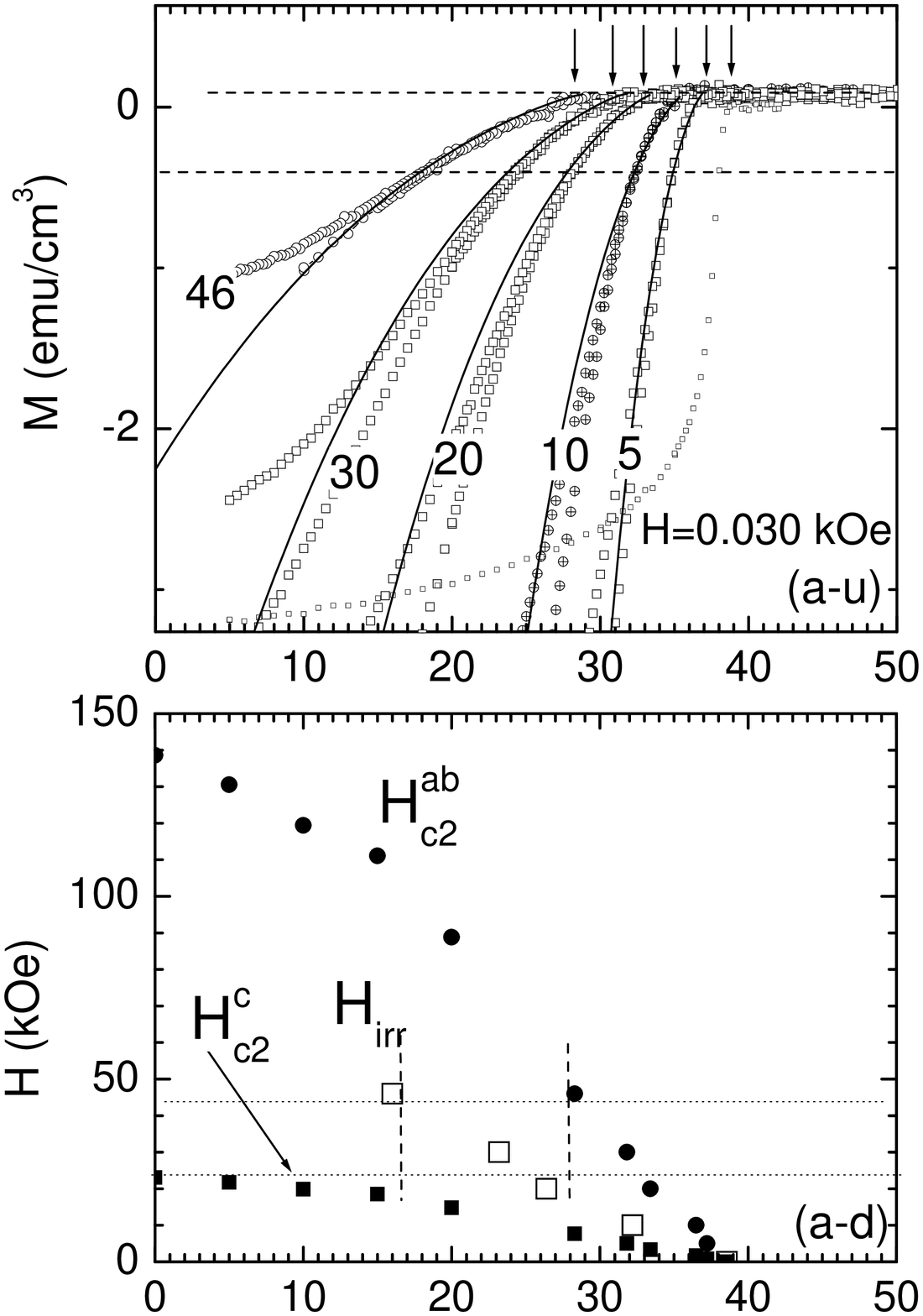}}
%
\centerline{\epsfxsize 7cm \epsfbox{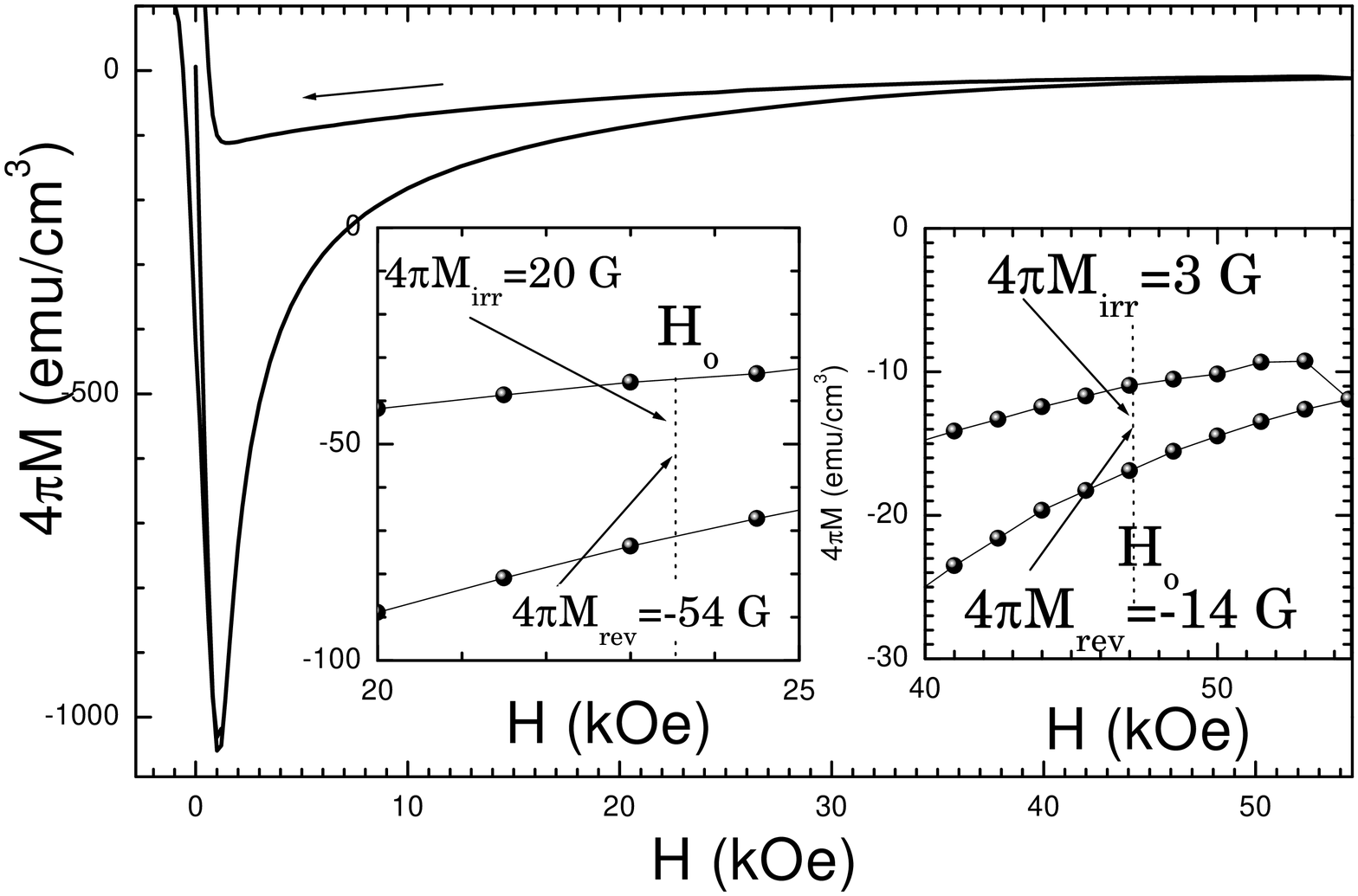}}
%

\caption{(a) (upper panel) Zero field and field cooling magnetic moment 
as a function of the temperature for $H=0.03,5,10,20,30$ and 46 kOe for the powder MgB$_2$ sample used
in the NMR measurements. The line through the experimental points is a simulation of the
reversible magnetic moment supposing an anisotropy $\gamma\approx 6$ (see main text).
The lower panel shows the phase diagram where, the filled circles and squares correspond
to $H_{c2}^{ab}$, $H_{c2}^{c}$ lines respectively. The open squares represent the irreversibility line.  
(b) The half of the magnetization 
loop at $T=5$ K. 
The insets show details in the region of $H=2.35$ and $H=4.7$ Tesla.}\label{fig1}%
\end{figure}

To investigate the mixed state and the magnetic irreversibility of MgB$_2$ 
we have employed thermomagnetic and isothermal magnetization measurements 
on a SQUID magnetometer. The upper panel of Figure \ref{fig1}a shows the 
zero field (ZFC) and field cooling (FC) magnetization curves for a
sample with randomly oriented grains, which was also used in the NMR
measurements. A marked feature is the observed curvature near
the onset of the transition, which has been recently attributed \cite
{Simon01} to the anisotropy of MgB$_2$. Specifically, the onset of the
diamagnetic signal occurs at the B$_{{\rm c2}}^{{\rm ab}}$ which varies 
with temperature as $B_{{\rm c2}}^{{\rm ab}}(T)=B_{{\rm c2}}^{{\rm ab}}(0)
(1-T/T_c)^{1.27}$. Since we examine a powder sample containing randomly 
oriented grains, we have applied the equation derived by Simon {\it et al} 
\cite{Simon01} for a uniaxial superconductor with the magnetization 
lying parallel to the external field, in order to fit the reversible part of the 
thermomagnetic curves. The solid line in Figure \ref{fig1}a shows the 
successful reproduction of the experimental data by using an 
anisotropy ratio $\gamma $$\sim 6$ and an average penetration depth: 
$\lambda =(\lambda _{{\rm ab}}^2\lambda _c)^{1/3}\approx 170$ nm, in 
agreement with the values in Ref.\cite{Simon01}. The lower panel of 
Figure \ref{fig1}a shows the phase diagram that is obtained from the 
magnetic measurements. This diagram includes the temperature variation 
of $B_{{\rm c2}}^c$ which is estimated from $B_{{\rm c2}}^c=B_{{\rm c2
}}^{{\rm ab}}/\gamma$. In this figure we also include the data of $H_{c2}^{ab}$
for $H>5.5$ T, from ref. \onlinecite{Simon01}. The irreversibility line is derived from the 
temperature where the ZFC and the FC branches are separated. Finally, 
Figure \ref{fig1}b shows the half of the isothermal hysteresis loop at 
$T=5$ K, which indicates that the irreversible magnetization is comparable
to the reversible one. This small irreversible magnetization, in the powder sample has been attributed 
to the surface barriers \cite{Pissas}. The two insets show details of the loop in the 
regions of 2.35 and 4.7 Tesla, which are used below to reproduce the 
magnetic field distribution in the vortex lattice.

\begin{figure}[tbp] \centering
\centerline{\epsfxsize 7cm \epsfbox{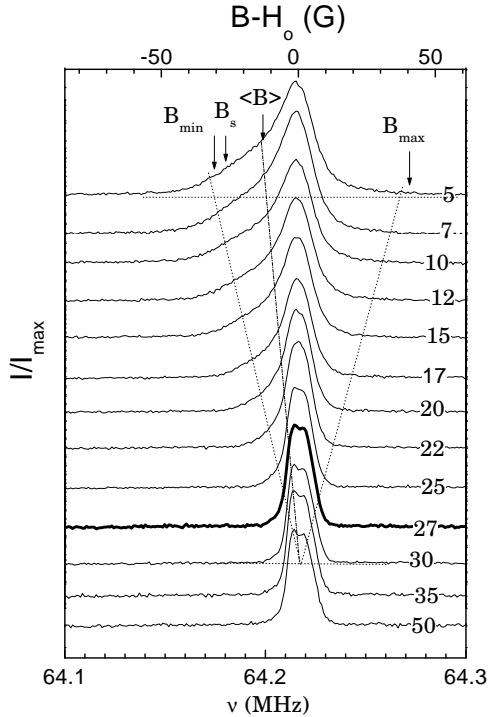}}
%

\caption{$^{11}$B NMR line shapes of the central transition in field $4.7$ 
Tesla for $5$ K$\leq T \leq 50$ K. The dotted lines correspond to $<B>$, 
$<B_{max}>$, and $<B_{min}>$.}\label{fig2}%
\end{figure}

$^{11}$B NMR line shape measurements were performed on two spectrometers
operating in external magnetic fields with $B_0=2.35$ and $4.7$ Tesla. The
spectra were obtained from the Fourier transformation of half of the echo,
following a typical $\pi /2$-$\tau $-$\pi $ spin-echo pulse sequence. In
both magnetic fields and in the normal state the length of the $\pi /2$
pulse was $t_p(\pi /2)\leq 2$ $\mu $sec, corresponding to a radiofrequency
(rf) irradiation field $B_1\geq 44$ Gauss. At room temperature the spectra
were found to exhibit the typical powder pattern for a nuclear spin $I=3/2$
in the presence of quadrupolar effects with an axially symmetric electric
field gradient, in agreement with previous works \cite{Gerashenko01,Jung01}.
The separation of the symmetric satellite lines gives a quadrupolar frequency 
\cite{Cohen57}: $\nu _Q=2\Delta \nu ^{(1)}=e^2qQ/2h\approx 0.836$ MHz. In
Figure \ref{fig2} we demonstrate the line shape of the central transition 
(-$\frac 12\longrightarrow \frac 12$) as a function of temperature in a
magnetic field of $4.7$ Tesla. The line shape in the normal state is
temperature independent and consistent with previous studies \cite
{Gerashenko01,Tou01,Jung01}. In addition, a second order quadrupolar split
of $\approx 6$ kHz is clearly observed. Considering that for $I=3/2$ the
second order quadrupolar split is given by \cite{Cohen57}: $\Delta \nu
^{(2)}=(25\nu _Q^2/144\nu _L)[I(I+1)-3/4]$, we obtain a $\nu _Q\approx 0.860$
MHz, in agreement with the value obtained from the separation of the
satellite peaks. 

Below $T_{{\rm c2}}$ ($\approx 27$ K in $4.7$ Tesla) the spectra start to 
broaden and under the irreversibility temperature an extra signal shows up 
as a pronounced shoulder in the low frequency part of the spectrum. It is 
worth noting that such an extra feature was not mentioned in previous NMR 
studies \cite{Gerashenko01,Jung01}. By further decreasing temperature the 
intensity of this shoulder increases and its location shifts to lower 
frequencies. This behavior is explicit to the magnetic field
distribution in the mixed state, as implied by the dotted lines.

A significant part of the signal remains unshifted at the frequency of the normal state NMR
signal. The unshifted part of the line shape may be explained if we consider
a distribution of $B_{{\rm c2}}$ caused predominately by anisotropy and not
by inhomogeneities, as the observed superconducting transition is extremely
sharp. Assuming that the upper critical field depends on the angle of the $c-
$axis in each crystallite then the angular dependence of the second critical
field is given by $B_{{\rm c2}}(\theta )=B_{{\rm c2}}^{{\rm ab}}(1+(\gamma
^2-1)\cos ^2\theta )^{-1/2}$. This equation shows that only crystallites
with $B_{{\rm c2}}(\theta )>B_{{\rm o}}$ would give a characteristic signal
of a type-II superconductor in the mixed state. If for a part of the grains $%
B_{{\rm o}}>B_{{\rm c2}}(\theta )$ then a NMR line shape would be observed,
which is the sum of spectra coming from crystallites in the normal state and
in the mixed state. Indeed the spectra under $B_{{\rm o}}=4.7$ Tesla show an
unshifted component which comes from the normal part of the sample,
revealing the presence of strong anisotropy.

\begin{figure}[tbp] \centering
\centerline{\epsfxsize 8cm \epsfbox{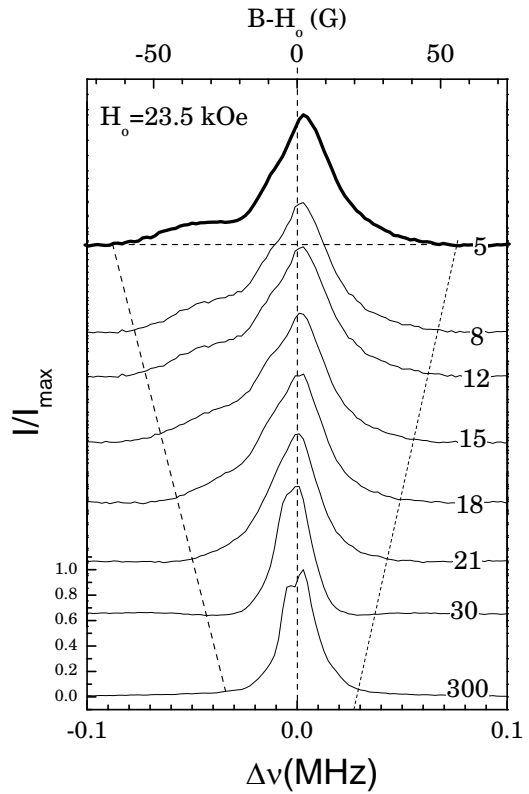}}
%

\caption{$^{11}$B NMR line shapes of the central transition in field $2.35$ 
Tesla, as a function of temperature.}%
\label{fig3}%
\end{figure}

The two signal components are more clearly resolved in Figure \ref{fig3},
which shows $^{11}$B NMR line shapes of the central transition in a magnetic
field $B_0=2.35$ Tesla. In this field the low frequency shoulder becomes
broader and shifts to lower frequencies in comparison to the spectra taken
in $4.7$ Tesla. This is expected if we consider that the magnetic field
distribution in the vortex lattice becomes less dense and exhibits stronger
field gradients in lower external magnetic fields. Since a significant part
of the signal intensity remains unshifted, then it can be argued that for a
part of the grains the $B_{c2}^c$ must be lower than $2.35$ Tesla.
Considering that magnetization measurements give $B_{c2}\approx 14$ Tesla,
and by assuming that $B_{c2}^{ab}>B_{c2}^c$ \cite{Lima01}, \cite{Simon01} an
anisotropy parameter $\gamma \geq 6$ is estimated.


Following Ref. \onlinecite{Brandt}, a rough estimation of the singular points (at London limit $\lambda>>\xi$) of
the field distribution at $T=5$ K and $H=4.7$ Tesla, gives: 
$B_{\max }-\left\langle B\right\rangle\approx  1.6\times 0.551 \Phi_{\rm o}/4\pi\lambda^2 \approx 50$ Gauss, 
$B_{\min }-\left\langle B\right\rangle\approx -0.6\times 0.551 \Phi_{\rm o}/4\pi\lambda^2 \approx -18$ 
Gauss and 
$B_s-\left\langle B\right\rangle\approx       -0.5\times 0.551 \Phi_{\rm o}/4\pi\lambda^2 \approx -15$ 
Gauss. In the above estimations we used 
the experimental value of the magnetic induction at $T=5$ K,
$\left\langle B\right\rangle=4\pi M_{rev}+H_{{\rm o}}\approx -14+H_{\rm o}$ Gauss
(see inset of Figure \ref{fig1}b) and $\lambda=170$ nm.

The estimated values of $B_{\max }$, $B_{\min}$ and $B_s$ are in good agreement 
with the characteristic points (see arrows in Fig. \ref{fig2}) of the NMR spectrum at $T=5$ K.
{\it These results indicate that the 
low frequency shoulder of the NMR spectra is produced by the magnetic field 
distribution of the vortex lattice}. 

The extension of the low field tail below the theoretical $B_{min}$%
, and the slightly shifted broad maximum in the experimental magnetic field
distribution might originate from small random pertubations by FL pinning or
by structural defects in the FLL, which lead to smearing of the ideal field
distribution. Deviation from the ideal distribution may be also produced by
shearing of the FLL, which leads to splitting of the singularity at $B_s$
and to the jump at $B_{\min }$, but leaves $B_{\max }$ approximately unchanged 
\cite{Brandt}. Smearing from random shearing would dominate, if
long-wavelength compression (flux density gradient) is not present, in
particular for small or large $B$ values, where the shear modulus $c_{66}$ $%
\longrightarrow 0$ for both $B/B_{c2}\longrightarrow 0$ and $%
B/B_{c2}\longrightarrow 1$ . On the other hand, the magnetic field
distribution is much more sensitive to fluctuations of the FLL density
(random compression) than to shear deformations for $\left\langle
B\right\rangle >>4\pi M$, if energetically favorable \cite{Comment}. In such
a case, even a small homogeneous compression of the FLL will shift rigidly
the magnetic field distribution density \cite{Brandt}, thus leading to
smearing of all van Hove singularities.
Finally, it should also be  noted that in anisotropic superconductors the field 
distribution deviates from that of an isotropic hexagonal vortex lattice 
\cite{Thie89}. 

In conclusion, $^{11}$B NMR line shape measurements on powder MgB$_2$ show
up two NMR signal components for $T<T_{c2}$: one coming from the vortex
lattice from grains in the mixed superconducting state, and another one from
grains that are still in the normal state as their orientation in respect to
the external magnetic field is such that $B_o>B_{{\rm c2}}(\theta )$. Our
measurements suggest a high anisotropy for the upper critical field with $%
\gamma \geq 6$, in agreement with recent CESR measurements \cite{Simon01}.
This experimental result changes the balance in favor of anisotropic
superconductivity in MgB$_2$, and should be taken into consideration in
theories trying to explain the superconducting state in this material.

%

\end{multicols}

\begin{references}
\bibitem{Comment}  Although we need higher energy to compress the FLL than
to shear it, the field distribution will be dominated by random compressions
when shear and compressional deformations will be of equal energy \protect\cite
{Brandt}.
\end{references}
\end{document}